\documentclass{article}
\usepackage{spconf,amsmath,graphicx}
\graphicspath{{figures/}{../figures/}}
\usepackage{makecell}
\usepackage{graphicx,hyperref}
\usepackage{hyperref}   
\usepackage{url}  
\hypersetup{
    colorlinks=true,
    linkcolor=blue,
    filecolor=magenta,      
    urlcolor=magenta,
}
 
\usepackage{booktabs}
\newcommand{\ra}[1]{\renewcommand{\arraystretch}{#1}}

\usepackage{enumitem}
\usepackage{amsmath}

\setlist[enumerate]{itemsep=0mm}

\title{SDCNet: Smoothed Dense-Convolution Network for Restoring Low-Dose Cerebral CT Perfusion}
\name{Peng Liu, Ruogu Fang}
\address{J. Crayton Pruitt Family Dept. of Biomedical Engineering, University of Florida}
\begin{document}
%
\maketitle
\begin{abstract}
With substantial public concerns on potential cancer risks and health hazards caused by the accumulated radiation exposure in medical imaging, reducing radiation dose in X-ray based medical imaging such as Computed Tomography Perfusion (CTP) has raised significant research interests. In this paper, we embrace the deep Convolutional Neural Networks (CNN) based approaches and introduce Smoothed Dense-Convolution Neural Network (SDCNet) to recover high-dose quality CTP images from low-dose ones. SDCNet is composed of sub-network blocks cascaded by skip-connections to infer the noise (differentials) from paired low/high-dose CT scans. SDCNet can effectively remove the noise in real low-dose CT scans and enhance the quality of medical images. We evaluate the proposed architecture on thousands of CT perfusion frames for both reconstructed image denoising and perfusion map quantification including cerebral blood flow (CBF) and cerebral blood volume (CBV). SDCNet achieves high performance in both visual and quantitative results with promising computational efficiency, comparing favorably with state-of-the-art approaches.  \textit{The code is available at \url{https://github.com/cswin/RC-Nets}}. 
\end{abstract}
\section{Introduction}
\label{sec:intro}

Computed tomography perfusion (CTP) is one of the most widely adopted imaging modalities for cerebrovascular disease diagnosis. 
However, CTP requires continuous scanning of a focused brain region to evaluate blood supplies, leading to growing concerns about the radiation damage to healthy cells, tissues, and organs~\cite{wintermark2010fda}.
Reducing the radiation exposure in CTP remains a critical and ongoing research direction.

Low-dose protocols have been developed and employed for reliable and low-risk CTP scanning. 
Decreasing the tube current is found to be the most effective approach to decrease the radiation exposure directly associated with CT imaging. However, it inevitably introduces significant image noise and artifacts due to the photon starvation artifacts~\cite{mori2013photon}. Plenty of image denoising methods have been widely investigated to maintain the image quality and accuracy, ranging from spatial filtering techniques ~\cite{ke2010multiscale,dabov2009bm3d} and iterative reconstruction methods~\cite{gu2014weighted} to deep convolutional neural network (CNN) based approaches, such as RED-Net~\cite{mao2016image} and DnCNN~\cite{zhang2017beyond}. Particularly, CNN-based strategy has achieved better overall performance than the traditional methods. 

In this work, we aim to study an effective, self-learning and scalable medical image denoising framework for reducing the radiation exposure in the imaging of cerebrovascular diseases. We investigate the effectiveness of the various structures of CNNs to overcome the noise challenge potentially existing in CTP images. More specifically, we find that larger kernel size ($>=7$) and larger number ($>=128$) of channels (output of feature maps) enable CNNs to learn prominent pixel-distribution features, which can promote the inference of noise-free images precisely. However, adopting larger kernel sizes also indicates larger border padding sizes ($>=3$) that result in hazy boundaries. This is because when a convolution window comes across the image boundary with large padding sizes, it yields blurred boundary due to the extrapolation based on incomplete information in the output. In addition, larger kernels and larger number of channels give rise to expensive computational costs; and, more importantly, large kernels lead to limited generalization ability, which is due to over-attention to large regions in pixel-distributions and overlooks the pixel representations extracted from relatively small reception fields.

\begin{figure*}[t]
\begin{center}
\vspace{-1em}
  \includegraphics[width=0.9\textwidth]{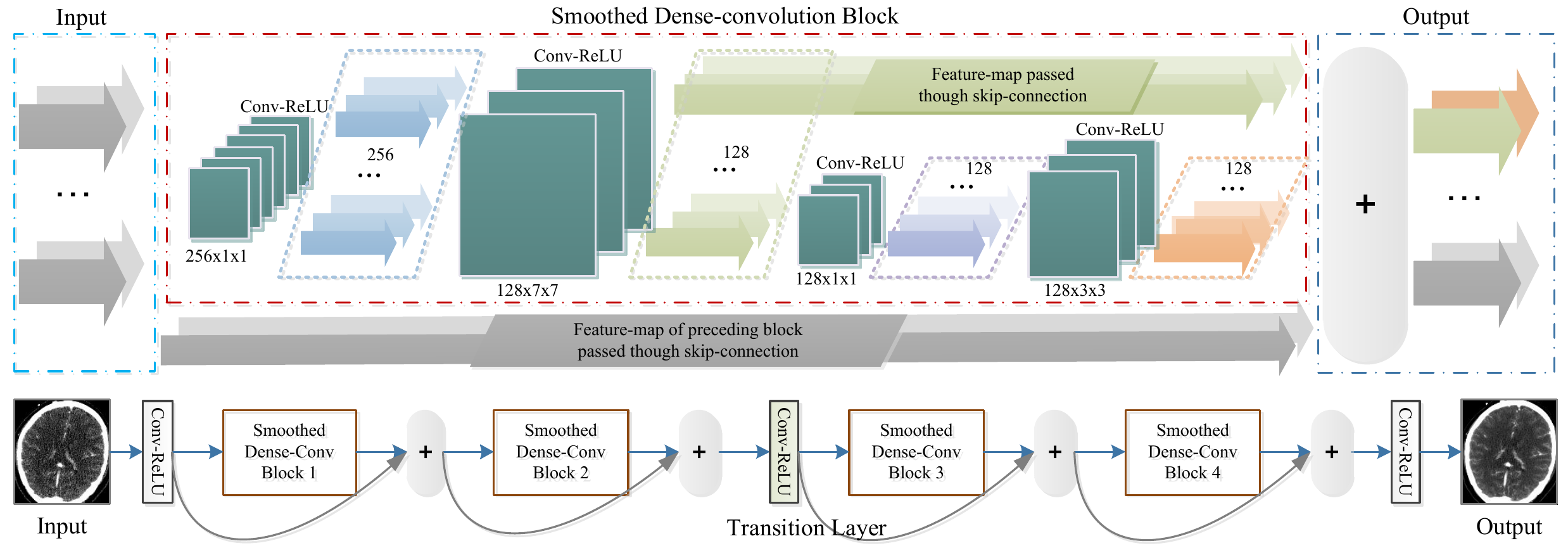}
  \caption{\small Top: A 4-layer smoothed dense-convolution (SDC) block with a massive of convolution computations (128x7x7) comprises two 1x1 convolution components for computation reduction and one 3x3 convolution module for alleviate the hazy border effect in the output images caused by employing large (7x7) convolution kernels . The number of dark-green blocks indicates the quantity of kernels in the current convolutional layers, and the size of dark-green blocks represents the area of convolution kernels and the density of convolution. The color arrows represent the quantity of feature-map outputs. Bottom: A deep SDCNet consisted of four SCN blocks. The layer between two adjacent blocks is referred to as transition layers and they change feature-map sizes via reducing convolution kernel number.}
\label{fig:SDCNet_stru}
\end{center}
\vspace{-2em}
\end{figure*}

From this new perspective, we propose a novel cascaded scalable architecture, called Smoothed Dense-Convolution Neural Network (SDCNet), which is composed of sub-network blocks (Fig.\ref{fig:SDCNet_stru}-top) cascaded by skip-connections. SDCNet is able to learn inference from the noise (differentials) between paired low/high-dose CT scan images in order to recover high-dose quality CTP images from the low-dose ones. SDCNet embedded with SDC blocks (Fig.~\ref{fig:SDCNet_stru}-top) can effectively remove the noise in real low-dose CT scan and enhance the quality of medical images under controllable convolution computational cost without the hazy border artifact. We evaluate SDCNet on thousands of CTP frames for both reconstructed image denoising and perfusion map quantification, including cerebral blood flow (CBF) and cerebral blood volume (CBV) maps. SDCNet has higher generalization ability and obtains outstanding performance in both visual and quantitative results with promising computational efficiency, comparing favorably with state-of-the-art approaches.

\section{MATERIALS AND METHODS} \label{sec:method}
\subsection{Dataset and Pre-Processing}\label{sbsec:dataset}
Our experiment was performed on real CT images scanned on human brains\footnote{Collaborate with Weil Cornell Medical College with patients' consent and IRB, HIPPA approvals}, and low/high-dose paired data of 26 aneurysmal subarachnoid hemorrhage  (aSAH) patients were collected. The primary parameters of the data acquisition were as follows: 80 kVp/160 mAs for high-dose scans, 80 kVp/20 mAs and 80 kVp/8 mAs for two different levels of low-dose scans. There are 27 frames for each slice and 16 slices in total for each patient in both low- and high-dose scans. Data prepossessing is described as below for deep CNN computation. 

To improve the robustness of the trained models for low-dose denoising, we preprocess the high / low dose frames before feeding them into the SDCNet. First, we remove non-brain regions from the whole-brain CT scans. Second, following the deep CNN-based image denoising models DnCNN~\cite{zhang2017beyond}, we crop brain regions into  $180\times180$ blocks. Last, we sample patches of size $40\times40$ from the  $180\times180$ blocks by a sliding window with a stride of 11, following the practice in DnCNN as well. Eventually we randomly select 505,000 low/high-dose paired CT scan patches for model training. 

\subsection{SDCNets Structure}
We build SDCNet by stacking multiple modularized sub-network blocks as shown in Fig.~\ref{fig:SDCNet_stru}-top. In this work, the structure of each sub-network block is designed with a hierarchical compound style, which starts with 256 $1\times1$ convolutional layer followed by 128 $7\times7$ convolutional layers. The network then separates into two branches. One branch starts with 128 $1\times1$ convolutional layers followed by  128 $3\times3$ convolutional layers. The other branch connects directly to the end of the block. The entire SDCNet starts with a transition layer composed of convolution and ReLU layers, followed by several sub-network blocks. For every block, the output of the two branches are element-wisely added to the output from the preceding block or a transition layer. This element-wise addition is primarily for reducing the number of output feature maps. We call this sub-network block as Smoothed Dense-Convolution (SDC) Block, where \emph{dense} indicates the large quantity of convolutional computations from large kernels and large channel numbers (aka number of output feature maps) in each block, and \emph{smoothed} indicates the combination of large and small kernels.

In our previous work~\cite{DBLP:journals/corr/LiuF17}, we empirically demonstrate that \emph{dense convolutions} via incorporating large kernel size ($>=7\times7$) (the area of convolution) with large channel number  ($>=128$) can effectively remove Gaussian noise from natural images. In this work, we focus on designing a more efficient dense-convolution network to address the realistic low-dose noise blending Gaussian and multiplicative noise in medical images, while tackling the hazy borders and the weak generalization issues. The effective reduction of the complex realistic low-dose noise relies on both dense convolutions and the depth of the networks (non-linear stacking layers). 

Specifically, given the limited paired of low/high-dose training data in practice, we adopt dense convolutions via large kernel sizes and large channel numbers to extract the most prominent features. This dense convolution network architecture aims to find the effective weights to infer the noise-free CT scans from the low-dose ones. On the other hand, to reduce the expensive computational cost of dense convolutions and ensure sufficient network depth for nonlinear inference, we build SDCNet with four dense-convolution blocks (Fig.~\ref{fig:SDCNet_stru}-bottom), where a $1\times1$ convolutional layer connecting adjacent blocks and  a $3\times3$ convolutional layer cascading $7\times7$ convolutional layers in each SDC block help to reduce the computational complexity, as shown in Fig.~\ref{fig:SDCNet_stru}-bottom.

The sizes of the convolutional kernels in each layer are chosen based on the specific purpose of each layer. In the proposed SDC block (Fig.~\ref{fig:SDCNet_stru}), the $1\times1$ convolutional layer is primarily for reducing the computation complexity. As we find that large convolution kernels lead to more boundary padding and result in hazy borders, we use $3\times3$ convolution kernels to extract the details around image borders and alleviated the impact of hazy boundary caused by large kernels.  

We also guarantee that dense-convolutional feature maps generated by large ($7\times7$) convolution kernels are the backbone of the SDCNet. To this end, we adopt two strategies. First, the $3\times3$ convolutional layers always follow the $7\times7$ convolutional layers to guarantee that images are convolved with the large kernels first. Second, each densely convolutional layer will directly pass its output feature maps to the end of the block through a skip connection. This refined design using a combination of large and small convolutional kernels ensures the $3\times3$ convolutional layer only addresses the hazy boundary issue without impacting the dense-convolutional feature maps for effective denoising.

Detailed layer information (from input to output) of the proposed SDCNet is listed below:
 
\begin{enumerate}[label=(\alph*)]
  \item \textit{Initial Layer}:  In this first layer, 128 convolution kernels of size $7\times7$ is followed by the leaky rectified nonlinearity unit (ReLU)~\cite{maas2013rectifier} for activation.
  \item \textit{Smoothed Dense-Conv Block 1, 2}: Each block contains 4 Conv-ReLU components, which has the following parameters for the convolutional layers: a 256 $1\times1$ convolution layers, followed by a 128 $7\times7$, a 128 $1\times1$, and a 128 $3\times3$ convolutional layers (Fig.~\ref{fig:SDCNet_stru}-top).
  \item \textit{Transition Layer (Conv-ReLU)}: This is also a Conv-ReLU component primarily for computational cost reduction via a smaller number of output feature maps (channel number), with 64 $7\times7$ convolutional layer.
  \item \textit{Smoothed Dense-Conv Block 3, 4}: Fewer channels are used compared to the Smoothed Dense-Conv Block 1 and 2, with a 128 $1\times1$ convolutional  layer, followed by a 64 $7\times7$, a 64 $1\times1$, and a 64 $3\times3$ convolutional layer (Fig.~\ref{fig:SDCNet_stru}-top).
  \item \textit{End Layer}: This end layer is used for compressing all features from the preceding blocks to one feature map as the final output with 1 $3\times3$ convolutional layer followed by ReLU. 
\end{enumerate}

The structural difference between DnCNN~\cite{zhang2017beyond} and our proposed network has two major points: (1) DnCNN adopts Bach Normalization (BN)~\cite{ioffe2015batch} but SDCNet does not. BN has a regularizing effect for accelerating training and enabling the use of higher learning rates, less careful
parameter initialization, and saturating nonlinearities. Without integrating with BN, we are more able to highlight SDCNet's characteristic. (2) DnCNN utilizes smaller size (64 $3\times3$) of space dimensions in all convolutional layers (except for the last layer 64 $1\times1$) than SDCNet.

\textbf{Differential learning:}

Rather than output noise-free images directly, SDCNet infers the noise map, which is the differential between low-dose and high-dose CT images. We call this process Differential Learning. This is an effective strategy for removing complex noise, which has been employed in DnCNN~\cite{zhang2017beyond}. Let us consider a low-dose noisy observation $y=x+n$. Here, $y$ and $x$ are a noisy low-dose CT observation and the corresponding latent clean high-dose CT image. $n$ represents the low-dose noise to be added to $x$. SDCNet aims to learn a mapping function \(T(y)\approx\displaystyle n\), and then it has \(x=y-T(y)\). In addition, to train this end-to-end differential
learning network by learning the convolution weights \(\Theta\),
we define the Mean Squared Error (MSE) between the noisy images (input \(y_{i}\))
and the real low-dose noise between the noisy image and the ground-truth images (\(y_{i}-x_{i}\)) as the loss function: 
\setlength{\belowdisplayskip}{0pt}  
\setlength{\abovedisplayskip}{0pt} 
\begin{equation}  \label{eq:1}
 l \left ( \Theta  \right )= \frac{1}{2N}\sum_{i=1}^{N}\left \| T\left ( y_{i};\Theta  \right ) - (y_{i} - x_{i})\right \|_F^2 
\end{equation}
and minimize it in $N$ training samples to learn the trainable weights \(\Theta\).

\section{EXPERIMENTS and RESULTS}
\label{sec:result}
SDCNet was trained on 505K low/high-dose paired image patches extracted from whole brain CT scans of 26 patients (Section~\ref{sbsec:dataset}). Stochastic gradient descent (SGD) algorithm was adopted to train the network with the step learning rate strategy. We set the basic learning rate at 0.1, momentum at 0.9, weight decay at $10^{-4}$, and clip gradient at 0.1. Batch size of 64 was used to balance the performance and training time. The network was trained for 50 epochs. 

Testing of SDCNet was performed using the slices from 12 individuals that are mutually exclusive with the training data. The image denoising performance was evaluated using two metrics: peak signal-to-noise ratio (PSNR) and Structural SIMilarity (SSIM)~\cite{wang2004image}.  Two perfusion maps (CBF and CBV)  are generated using the following steps for performance evaluation: (1) Remove noise in low-dose CT images using the trained SDCNet model. (2) Converting the output of the SDCNet to DICOM files by merging the meta data obtained from original low-dose DICOM data with the denoised images. (3) Processing the new DICOM files with  the bSVD algorithm~\cite{wittsack2008ct} to generate perfusion maps: CBF and CBV. 

\begin{figure}[t]

\begin{center}
  \includegraphics[width=\linewidth]{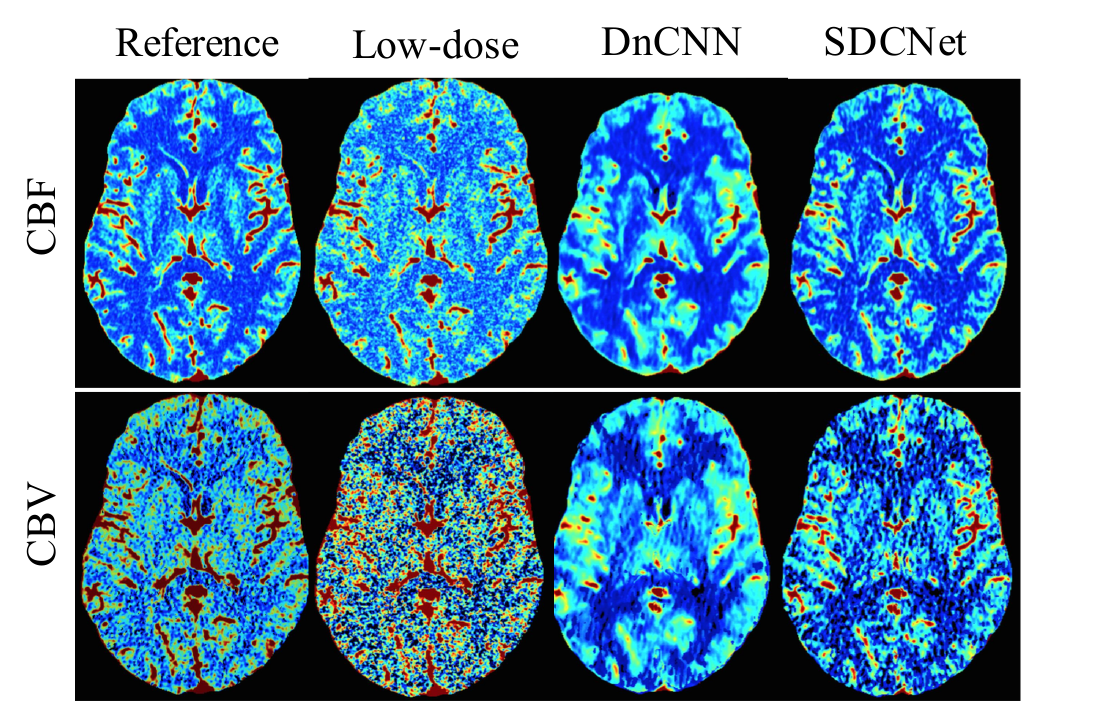}
  \caption{\small Comparison of perfusion maps produced respectively via DnCNN~\cite{zhang2017beyond} and SDCNet based on 20mAs low-dose CT scan data.}
\label{fig:persusion}
\end{center}
\vspace{-2em}
\end{figure}

We compared the image denoising performance with deep CNN-based DnCNN~\cite{zhang2017beyond}, a state-of-the-art image denoising algorithm. Evaluation metrics used are average PSNR, SSIM and the computation time. As shown in Table~\ref{tab:result}, SDCNet obtained large performance gains over DnCNN on both image quality metrics. We also compared the perfusion maps generated the denoised CTP sequences. As shown in Figure~\ref{fig:persusion}, while DnCNN over-smooothed both CBF and CBV maps, our proposed SDCNet was able to restore the high-fidality blood flow patterns by preserving the vessel details and the white-matter gray-matter differences. 

\begin{table}[t]
 \centering 
\caption{\small The average results of PSNR (dB) / SSIM / Run Time (seconds) of different methods (DnCNN and the proposed SDCNet) on 1,100 real low-dose (20mAs) whole-brain CT scan frames from 10 different individuals and 110 frames randomly selected from each individual. Bold font indicates the best performance.}
\ra{1}
\resizebox{1\columnwidth}{!}{
\begin{tabular}{@{}rrrrcrrrcrrr@{}}\toprule
& \multicolumn{2}{c}{Low-dose} & \phantom{abc}& \multicolumn{3}{c}{DnCNN\cite{zhang2017beyond}} &
\phantom{abc} & \multicolumn{3}{c}{SDCNet}\\
\cmidrule{2-3} \cmidrule{5-7} \cmidrule{9-11}
& PSNR      & SSIM         && PSNR     & SSIM   & time          && PSNR          & SSIM          & time         \\  \midrule
& 18.23     & 0.4903       && 33.83         & 0.7422        & \textbf{30.14}        && \textbf{34.69}   &  \textbf{0.7833}  & 35.17 \\
\bottomrule
\end{tabular}
\label{tab:result}
}

\vspace{-1em}
\end{table}

\section{CONCLUSION}\label{sec:conclu}
A novel deep learning-based SDCNet was proposed and applied to restoring low-dose cerebral CT perfusion scans. The proposed network consists of a combination of large and small kernels with dense convolutions as the backbone of the architecture. With limited number of real paired low/high-dose brain CT scan data, SDCNet can be trained to infer the difference (noise) between the paired low-dose and high-dose CT scans effectively and recover high-dose quality cerebral CT perfusion maps from the low-dose ones.

In addition to accurate perfusion images recovery, the proposed SDC blocks in the proposed framework are scalable. This smoothed block is featured by a combination of large and small convolution kernels. On one hand, large kernels extracts prominent features efficiently via increased kernel size and channel numbers. On the other hand, the introduction of smalle kernels help to balance the hazy boundary artifacts. Moreover, it can decrease the model parameters via interpolating $1\times1$ convolutional layers to avoid overfitting and expensive computation cost. Finally, high-quality cerebral perfusion maps generated from low-dose CTP sequences demonstrated the promising potential of aiding medical diagnosis of brain diseases at significantly lower radiation exposure and potential long-term health risks. In our future work, we will use the proposed network to predict the high-dose perfusion maps (CBF, CBV, etc.) directly from the low-dose sequential CTP images, skipping the intermediate step of recovering high-dose quality CTP images.

\bibliographystyle{ieee}
\small{
    \bibliography{refs.bib}
}

\end{document}